\documentclass[3p,12pt,times]{elsarticle}

\usepackage{amssymb}

\journal{Renewable Energy}

\begin{document}

\begin{frontmatter}



\title{On the comparison of energy sources: feasibility of radio frequency and ambient light harvesting.}


\author[unm-math]{Alexander O. Korotkevich\corref{kao}}
\ead{alexkor@math.unm.edu}
\cortext[kao]{Corresponding author}

\author[unm-math]{Zhanna S. Galochkina}
\ead{zhanna@unm.edu}

\author[unm-ece]{Olga Lavrova}
\ead{olavrova@ece.unm.edu}

\author[sb]{Evangelos Coutsias}
\ead{vageli@math.unm.edu}

\address[unm-math]{Department of Mathematics \& Statistics,
University of New Mexico,
MSC01 1115, 1 University of New Mexico,
Albuquerque, NM 87131-0001, USA}

\address[unm-ece]{Department of Electrical \& Computer Engineering,
University of New Mexico,
MSC01 1100, 1 University of New Mexico,
Albuquerque, NM 87131-0001, USA}

\address[sb]{Department of Applied Mathematics \& Statistics,
Stony Brook University, Stony Brook, NY 11794-3600, USA}

\begin{abstract}
With growing interest in multi source energy harvesting including
integrated microchips
we propose a comparison
of radio frequency (RF) and solar energy sources in a typical city.
Harvesting devices for RF and solar energy will be competing for space
of a compact micro or nano device as well as for orientation with respect to the energy source.
This is why it is important to investigate importance of every source of energy
and make a decision whether it will be worthwhile to include such harvesters.
We considered theoretically possible irradiance by RF signal in different situations,
typical for the modern urban environment and compared it with ambient solar energy
sources available through the night, including moon light.

Our estimations show that solar light energy dominates by far margin
practically all the time, even during the night, if there is a full moon
in the absence of clouds. At the same time, in the closed compartments or at
the new moon RF harvesting can be beneficial as a source of ``free'' energy.
\end{abstract}


\end{frontmatter}


\section{Introduction}
Consideration of alternative and ``free'' energy sources
as a power source for different devices, including microelectronics, is an important and popular topic~\cite{TIharvesting,DigiKeyharvesting,SLharvest}.
One of the most promising directions is energy harvesting -- utilization of energy available
as a background radiation noise: solar radiation, light from the lamps in the building, Wi-Fi or cell phone radio frequency (RF)
radiation etc.
Let us estimate the potential for energy harvesting from different sources. 
We shall not consider vibration or thermal sources,
because they greatly depend on the environment and may be highly intermittent. 
We thus concentrate on solar energy, moonlight, ambient city light, and RF energy,
because these types of energy are available in any modern city. 
For example, if we consider a photovoltaic (PV) panel augmented with
RF-antennas and conversion microcircuit integrated into the horizontal surfaces of an automobile 
(roof, hood, tint film on windows etc.) to power different micro-sensors it is useful to know how much energy we can
potentially harvest from different sources, in order to make a decision already on the design stage
about the cost effectiveness of integrating of RF antennas and conversion microcircuits,
which usually include Schottky diode, into the PV-devices. While for the usual micro-controllers (e.g.~\cite{TIharvesting})
there is no strict limitation in space and addition of energy conversion circuit will not change the price of
device significantly, for micro and nano scale devices the space limitation can be already crucial.

Our estimations show that in most of situations solar energy will be dominant by at least order of magnitude.
Even during the night moon light
can provide more energy than RF-radiation, if we do not consider cases when the Moon is absent. During dark nights
or in environments with blocked solar radiation RF energy source can be valuable.

The plan of the paper is as follows: first we consider different types of energy sources and estimate irradiance generated
by them -- power per unit area. After that, taking into account conversion rate 
of available harvesting devices, we discuss feasibility of different sources in the conclusion.

\section{Solar energy}
As a reference point we used data about surface irradiance collected by 
the Photovoltaic Systems Evaluation Laboratory in
Sandia National Laboratories for latitude 35.111 N, longitude 106.61 W (Albuquerque, New Mexico, USA) during 
a twenty four years period
(from 1989 to 2012 year)~\cite{24yearSummarySolarSandia}. Data for different locations can be found at NASA Surface meteorology and Solar Energy web-site~\cite{NASAsolar}. Global horizontal 
irradiance (GHI, the total amount of shortwave radiation received from above 
by a horizontal surface) averaged over a twenty four years period yields
the value
\begin{equation}
\label{I_Solar}
I_{Solar} = 1.1\times 10^{3} W/m^2.
\end{equation}
This value can vary due to different factors, like seasonal monsoon rains, 
volcanic eruptions (Mt. Pinatubo, Philippines, erupted 6/91, 
reduced solar resource in late 1991, 1992, early 1993; 
maximum decrease in GHI was about 10\% and was observed in
1992~\cite{24yearSummarySolarSandia}). Nevertheless for order of magnitude estimation such fluctuations 
are definitely beyond the required accuracy.

Conversion efficiency also affects all sources of energy so we have to take 
it into account. For PV cells
conversion rate depends on the material and type of the cell. Widely used cells based on crystalline silicon with
single p-n junction have a theoretical limiting power efficiency of 33.7\% (Shockley-Queisser limit~\cite{SQ1961}) and
current crystalline silicon based PV cells are approaching this limit, while state-of-the-art multijunction samples
with concentrators can yield up to 44.4\% efficiency~\cite{SharpRecord}. 
Taking into account the packing ratio (not the whole surface of the PVC panel is occupied by a converting element)
of real PV devices
conversion efficiencies between 17\% and 20\% are typical for commercially available Si-based solar cells.

\section{Light from the Full Moon}
\subsection{Theoretical estimation}
During the night the Moon can be a significant source of light depending on phase of the Moon cycle. 
It is useful to compare irradiance from this
source of energy with other options.
Let us estimate irradiance due to full moon in reasonably good conditions. We shall calculate an attenuation factor similar to one
introduced in~\cite{LB2000} (the factor was not calculated in the paper and proposed formula gave result several orders of magnitude lower even for wrong albedo value which was order of magnitude higher than the real one due to derivation error).

Earth and Moon are on average at approximately the same distance from the Sun, so for our rough estimation we can consider the irradiance from the Sun as the same one.
This estimation is confirmed by measurements~\cite{NASAMoonAlbedo}, if we shall take into account albedo of Earth atmosphere which is approximately 30\%.
Moon's mean radius is $R_{M}\simeq 1738\, km$~\cite{NASAMoon}, which is significantly smaller than Earth's mean radius $R_{E} \simeq 6371\, km$~\cite{NASAEarth}
resulting in smaller amount of energy obtained by Moon from Sun, which is proportional to the surface area. This means that the reduction coefficient
for energy obtained by Moon from Sun will be $R_M^2/R_E^2 \simeq 0.074$. Moon albedo is $\alpha \simeq 0.12$~\cite{NASAMoonAlbedo} (here we neglect the fact
that Moon albedo depends on the light incident angle, resulting in stronger reflection if the source of light is behind the observer~\cite{BHW1996}, which is
typical for the full Moon), which means that only 12\% of incoming radiation
is scattered in all directions over the surface (i.e. in upper hemisphere). In order to figure out what fraction of this energy reach
the Earth we need to compare the solid angle of the whole hemisphere of scattering (which is $2\pi\,\mathrm{Sr}$) and solid angle of the Earth from Moon. Average distance
from Earth to Moon is $L_{EM} \simeq 384400\, km$~\cite{NASAMoon}, which give us the solid angle
$\Omega = 2\pi(1 - L_{EM}/\sqrt{R_E^2 + L_{EM}^2}) \simeq 0.86\times 10^{-3}\,\mathrm{Sr}$.
The resulting attenuation coefficient for the Sun radiation can be estimated as follows:
\begin{equation}
C_{Moon} = \left(\frac{R_M^2}{R_E^2}\right)\alpha\left(\frac{\Omega}{2\pi}\right)\simeq 1.2\times 10^{-6}.
\end{equation}
Taking into account this attenuation coefficient one can get the following estimation for the irradiance from the Moon light
\begin{equation}
\label{Moon_theory}
I_{Moon}^{Theory} = C_{Moon} I_{Solar} \simeq 1.3\times 10^{-3} W/m^2.
\end{equation}
Here we supposed that moonlight attenuation in the Earth atmosphere is approximately the same as for the sunlight.

It should be noted that this number decrease significantly with different 
phases of the Moon, resulting in its complete extinction when there is a 
new Moon or if Moon is under horizon. Some estimations of the availability of this source of power is given in Conclusion. It follows from these calculations that Moon is present at dark times only 25\% of the time.

Because we consider the conversion of the moonlight by PV cells 
we shall use the same
conversion rate about 20\% as for the sunlight conversion. For our rough analysis we can neglect difference in the spectra.

\subsection{Measurements}
Following~\cite{BM1969}, illumination from the full
moon (at $60\,^{\circ}$ elevation angle) can achieve $0.7 \ lux$ which roughly corresponds to
\begin{equation}
\label{I_Moon}
I_{Moon} = 1.0\times 10^{-3} W/m^2.
\end{equation}
This value is in good correspondence with our estimation~(\ref{Moon_theory}). For convenience and in order to be on the safe side when comparing with
different sources later in the article we shall use this value for the irradiance from the Moon.

We have performed a measurement of harvested moonlight power at full moon conditions at the same location (Albuquerque, NM, USA). No visible clouds were present during the measurement night. Two of standard 6-inch mono-crystalline Si solar cells, each of surface area of $0.0235 m^2$. Both cells were connected in series, forming a miniature PV panel of total area of $0.047 m^2$. Each of these solar cells had nominal STC  (Standard Test conditions) conversion efficiency of 21\%, as reported by the cell manufacturer (Schott Solar). Under a full moon conditions, and at normal incidence to moonlight, a total current of $1.5\times 10^{-5} A$ at voltage $1.2 V$ was measured from this miniature panel, which gives $1.8\times 10^{-5} Watt$. This measurement translates into $\simeq 3.8\times 10^{-4} W/m^2$ produced by the panel, or, taking into account 21\% conversion efficiency,  $1.8\times 10^{-3} W/m^2$ of moon power available for harvesting  under realistic conditions. This number is slightly higher compare to the value estimated above -- but this is a realistic expectation due to relatively high elevation of the measuring site (Albuquerque, NM is situated at the approximately $1.6\; km$ above the sea level), good atmosphere transparency, and relatively low air pollution.

\section{Interior lighting}
Interior lighting  can also be a source of energy inside the building. In order to evaluate possibility of using interior lighting for energy harvesting, we performed a similar measurement of a harvested interior lighting energy with the same miniature Si PV panel. In a typical US office environment with fluorescent lighting, a total of $4.5 mWatt$ was measured. This measurement translates into approximately $0.1 W/m^2$ produced by the panel, or, taking into account 20\% conversion efficiency, $0.5 W/m^2$ of interior lighting available for energy harvesting purposes. Of course, the actual amount of interior light will vary greatly from building to building, and office to office. Inside the building if we consider inhabited areas the artificial light sources are present and can be sources of energy.
We have to emphasize, that for the currently most popular indoor office lighting systems one should use specific
types of PV cells. We should note that interior lighting spectra (typically from fluorescent lighting) is significantly different from outdoors reference AM1.5 (Air Mass) spectrum -- with fluorescent emission having significantly more energy concentrated in longer wavelengths (visible to IR regions) and significantly less energy concentrated in shorter wavelength. Therefore, previously mentioned Si solar cells are no longer an efficient indoor lighting harvester material. Other semiconductors, such as CdTe, CdS or more expensive GaSb~\cite{GaSb} are better match (due to lower semiconductor band gaps). CdTe have been used historically for multiple decades powering up small calculators -- therefore proving the ideas of feasible energy harvesting even indoors. Future materials, such as GaSb~\cite{GaSb} may improve the utilization efficiency of such indoor energy harvesting even further. So if the harvester has to be situated indoors these types of PV cells have to be considered.

\section{Radio Frequency energy}
In a modern city we are surrounded by different sources of radio frequency energy. E.\,g., Wi-Fi access point,
cell phone towers, medical equipment, home radiophones, etc. Usually frequency range of electromagnetic radiation
from these sources is about $2 \ GHz$ and, typically, not higher that $5.8 \ GHz$. Experiments on microwave power transmission
had shown feasibility of this technology already with pioneering works by Brown in 1960s~\cite{Brown1984}
and 1990s~\cite{BE1992}.
For conversion of the microwave signal into DC current a rectifying antenna (or rectenna) is used. Efficiency
of such devices, usually based on Schottky diode, can be as high as 90\% for relatively high power levels (90.6\%
at frequency $2.45 \ GHz$ and power $8 \ W$~\cite{Ratheon1977}) and decreases 
slightly for lower power levels (82\%
at an input power level of $50 \ mW$ and frequency $5.8 \ GHz$~\cite{McSFC1998}) with further decrease of efficiency
at even lower power levels (15.7\% and 42.1\% were obtained for input available power levels of $0.01 \ mW$ and $0.10 \ mW$ respectively at $2.45 \ GHz$~\cite{VGCV2010}).

Let us estimate irradiance levels from different RF sources. Following FCC Regulation for Wi-Fi equipment in $2.4 \ GHz$ band,
Wi-Fi access point with omnidirectional antenna of less than $6 \ dBi$ gain has to have equivalent isotropically radiated power (EIRP) less than $1 \ W$~\cite{FCC_WIFI}. In order to estimate irradiance let us divide the power by the area of a sphere of the radius which is
equal to the distance to the radiation emitter. For example, at the distance $100 \ m$ from the access point (which is not
really that far, because we usually do not have a Wi-Fi access point on an open field, but in some building resulting in multiple reflections on the way of propagation)
scaling factor will be $1/(4\pi 100^2) m^{-2}$ which yields irradiance for Wi-Fi:
\begin{equation}
\label{I_WiFi}
I_{WiFi} = 0.01\times 10^{-3} W/m^2.
\end{equation}
Here we used very optimistic estimation for the power of transmitter. In reality EIRP is about order of magnitude lower. In a recent work~\cite{Foster2007} 356 measurements were performed at 55 sites of different type (including residential areas, commercial spaces, health care providers, educational institutions, and other public areas) in four countries (USA, France, Germany, and Sweden) in wide range of frequencies (75MHz -- 3GHz). Power density was practically always lower than $10^{-5} W/m^2$ which is in agreement with our estimations. Taking into account that in the mentioned work distances to the sources of radiation were typically much smaller (several meters or so) we can safely state that our estimations of RF energy sources give us upper limit for a typical situation.

Now let us consider radiation from the cell tower, which is situated at a distance of $300 \ m$ (conservative estimation, due to buildings with resulting signal reflections this distance is higher). For the distance estimation a map of cell towers in Albuquerque, NM~\cite{CellTowers} was used. Following information from FCC~\cite{FCC_CELLTOWERS}
typical EIRP in the majority cases of urban or suburban cell towers is $100 \ W$. Taking into account scaling factor
$1/(4\pi 300^2) m^{-2}$ we get irradiance from the cell tower:
\begin{equation}
\label{I_Cell}
I_{Cell} = 0.1\times 10^{-3} W/m^2.
\end{equation}

Other sources of RF energy can be considered. For instance, smart power meters~\cite{FT2013}. At the same time most of these
sources are either very rarely active or significantly less powerful, than estimates for radiation from cell towers
(typical irradiance for most of the sources is not higher than $0.01\times 10^{-3} W/m^2$, with exclusion of mobile handsets,
which has to be situated very close to the harvesting device).
For a good review on different sources of RF energy we would recommend recent article~\cite{FM2013}.

Thus, we shall limit ourselves with considered sources of RF energy. As one of the most prominent alternatives in the rural
areas one should consider high voltage power lines, due to relatively
high irradiance (about $2\div 4 \ W/m^2$) and relatively slow decay of irradiance with distance (due to cylindrical geometry
decay is proportional to the distance, instead of squared distance like in previous examples). Although, such sources of energy
are very rare in urban areas.

\begin{table}[hb!]
\center
\begin{tabular}{|ccccc|}
\hline
Solar & Interior lighting & Moon & Cell Tower & WiFi  \\
\hline
$1.1 \times 10^3$ & $0.5$ & $1.0 \times 10^{-3}$ & $0.1 \times 10^{-3}$ & $0.01 \times 10^{-3}$  \\
\hline
\end{tabular}
\caption{\label{tab:comp} Comparison of different sources of energy (order of magnitude estimations)  ($W/m^2$).}
\end{table}

\section{Discussion and Conclusion}
If we consider a rectenna with effective area about $1 \ m^2$ (which is an 
optimistic estimation), then taking into account
dependence of conversion efficiency on the power we get 40\% conversion rate for RF energy as a top bound.
Such value is of the same order as an average conversion rate for PVCs. As a result we can compare just pure irradiances.
As one can see, even during the night, if we have full moon, irradiance due to the moonlight (\ref{I_Moon}) will be
at least an order of magnitude higher than the one due to RF-signal from 
a cell tower (\ref{I_Cell}), and two orders of magnitude
if we compare with Wi-Fi (\ref{I_WiFi}). From the other side, if we are really close to the access point, the signal can be
several times stronger. During the day, even if we have a full sun eclipse or if we hide in the shade,
irradiance due to sunlight - direct or diffuse - will be by several orders 
of magnitude higher than any other ``free'' sources. Indoor irradiance from the usual office lighting also surpasses RF sources by several orders of magnitude.

Also we should take into account that moon visibility during nights is approximately 25\% even if consider not the full moon. Clearly the Sun and Moon are each below the horizon 50\% of the time on average, and since their
motions are incommensurate, they will be alternatively in the sky together or apart in equal times.
Therefore, approximately a quarter of the time there is no strong light source in the sky (for direct computations of Moon visibility one can use a Matlab script~\cite{MatlabScript}). Thus, RF-sources have to be taken into account when one is developing harvesting device which will operate in rural areas far from the ambient light sources,
like the usual city street lights.

As a result we can formulate a simple recommendation: during the day sun is the main source of energy, during the night, if there are clouds or a new moon RF-sources may be beneficial, although in urban areas ambient light can provide even more energy. If there is a full or close to full moon -- PV cells should be used
in order to harvest energy from the moonlight. This recommendations allow to make a decision about augmentation of the PV cells
with rectennas. Usually rectennas are based on Schottky diodes, which are relatively expensive piece of equipment, so unless a new substantially cheaper
technology is introduced rectennas
should be integrated into PV cells only if the harvesting device is going to work in dark environments or close to cell towers, Wi-Fi access points and other sources of RF radiation. We would like to emphasize that our estimations were performed
for typical situations in urban areas and they have to be adjusted in the case of close proximity of e.g. high voltage power transmission lines, powerful TV and radio transmitters, or other significant sources of RF radiation.

\section{Acknowledgments}
This work was partially supported by the NSF grant CHE 1231046.



\end{document}